\documentclass{iaus}
\usepackage{graphicx}

\title[SS\,433 and ultraluminous X-ray sources] 
{The supercritical accretion disk in SS\,433 and ultraluminous X-ray sources}

\author[S. Fabrika, P. Abolmasov and S. Karpov]   
{S. Fabrika, P. Abolmasov \and S. Karpov}

\affiliation{Special Astrophysical Observatory of the Russian AS, Nizhnij
             Arkhyz 369167,  Russia}

\pubyear{2006}
\volume{238}  
\pagerange{001--999}
\date{October 15, 2006 and in revised form ???}
\jname{Black Holes: from Stars to Galaxies -- across the Range of Masses}
\editors{V. Karas \& G. Matt, eds.}
\begin{document}

\maketitle

\begin{abstract}
SS\,433 is the only known persistent supercritical accretor, it may
be very important for understanding ultraluminous X-ray sources (ULXs)
located in external galaxies. We describe main properties of the SS\,433
supercritical accretion disk and jets. Basing on observational data of SS\,433
and published 2D simulations of supercritical accretion disks we estimate
parameters of the funnel in the disk/wind of SS\,433. We argue that the 
UV radiation of the SS\,433 disk
($\sim 50000$ K, $\sim 10^{40}$ erg/s) is roughly isotropic, but X-ray
radiation ($\sim 10^7$ K, $\sim 10^{40}$ erg/s) of the funnel is midly
anisotropic. A face-on SS\,433 object has to be ultraluminous in X-rays
($10^{40-41}$ erg/s). Typical time-scales of the funnel flux
variability are estimated. Shallow and very broad (0.1-0.3c) and blue-shifted
absorption lines are expected in the funnel X-ray spectrum.
\end{abstract}

\firstsection 
\section{Introduction}

The main properties of the ultraluminous X-ray sources (ULXs) are their
extremely high luminosities ($10^{39-41}\,erg/s$), diversity of X-ray spectra,
strong variability, their connection with star-forming regions and surrounding
nebulae. Here we continue to develop the idea that the galactic supercritical accretor
SS\,433 intrinsically is very bright X-ray source and it is a prototype of ULXs
in external galaxies (\cite{Katz87, FaMe01, Kingetal01, Begelman06, Pout06}).
This means that ULXs are supercritical accretion disks in close binaries with stellar 
mass black holes or microquasars. We discuss possible properties of the funnel 
in the supercritical accretion disk of SS\,433 to predict X-ray spectral 
features and temporal behaviour of the funnel in "face-on SS\,433" stars in application
to ULXs.

\section{Properties of SS\,433}

The main difference between SS\,433 and other known X-ray binaries is
highly supercritical and persistent mass accretion rate
($\dot M_a \ge 10^{-4}\,M_{\odot}/yr$) onto the relativistic star, a probable
black hole ($M\sim 10 M_{\odot}$), which has led to the formation of a
supercritical accretion disk and the relativistic jets. SS\,433 properties
were reviewed recently by \cite{Fab04}. 

A total observed luminosity of SS\,433 is $L_{bol} \sim 10^{40}\,erg/s$. 
Practically all the energy is realised near the black hole.
A temperature of the source
is $T = (5-7)\cdot10^4\,K$, when the accretion disk is the most open to the
observer and it is $T \sim 2\cdot10^4\,K$, when it is observed edge-on 
(\cite{dolan97}). If the source SED is represented by a single black-body source,
its size is $\sim 10^{12}\,cm$. 
The line of sight wind velocity varies from $\sim 1500\,km/s$
(the most open disk) to $\sim 100\,km/s$ (the disk is observed edge on).
One may adopt the value $\sim 2000\,km/s$ for the wind velocity closer to the jets.   
The mass loss rate in the wind is $\dot M_w \sim 10^{-4}\,M_{\odot}/yr$.
The optical jets ($\sim 10^{15}\,cm$) consist of dense small gas clouds,
the X-ray jets ($\sim 10^{12}\,cm$) consist of hot ($T \sim 10^8\,K$) cooling plasma.
Both the optical and X-ray jets are well collimated ($\sim 1^{\circ}$).
SS\,433 X-ray luminosity (the cooling X-ray jets) is $\sim 10^4$ times less than
bolometric luminosity, however kinetic luminosity of the jets is very high, 
$L_{k} \sim 10^{39}\,erg/s$, the jet mass loss rate is
$\dot M_j \sim 5 \cdot 10^{-7}\,M_\odot/$y

The jets have to be formed in a funnel in the disk or the disk wind.
Supercritical accretion disks simulations (\cite{Egetal85, Ohsuga05, 
Okudaetal05}) show that a wide funnel ($\theta_f \approx 20^{\circ}-25^{\circ}$) 
is formed close to the black hole. Convection is important in the inner 
accretion disk. Recent hydrodynamic simulations of super-Eddington 
radiation pressure-dominated disks with photon trapping (\cite{Ohsuga05})
have confirmed the importance of advective flows in the most inner parts of 
the disks. The mass accretion rate into the black hole a few times exceeds 
the mass loss rate in the disk wind. One may adopt a rough estimate of the 
accretion rate in the case of SS\,433 (or the same for the gas supply rate 
in the accretion disk), $\dot M_a \sim 3 \cdot 10^{-4}\,M_{\odot}/yr$.

\section{The funnel in the supercritical accretion disk}

A critical luminosity for a $10\,M_{\odot}$ black hole is 
$L_{edd} \sim 1.5 \cdot 10^{39}\,erg/s$ and corresponding mass accretion rate
$\dot M_{edd} \sim 3 \cdot 10^{-7} \,M_{\odot}/yr$. 
There are three factors (\cite{FK06}) increasing observed face-on luminosity
of a supercritical accretion disk. (1) inside the spherization radius 
the disk is locally Eddington (\cite{SS73}), which gives logarithmic factor
$(1 + \ln(\dot M_a / \dot M_{edd})) \sim 8$; (2) the doppler boosting factor is 
($\beta=V_j/c=0.26$ in SS\,433) is $1 / (1-\beta)^{2+\alpha} \sim 2.5$, where
$\alpha$ is spectral index; and (3) the geometrical funneling 
($\theta_f \sim 25^{\circ}$) is $\Omega_f/2\pi \sim 10$. 
Thus, one may expect an observed face-on luminosity of such supercritical disk 
$L_{edd} \sim (2-3) \cdot 10^{41}\,erg/s$. 
On the other hand, if one adopts for the funnel luminosity, that it 
is about the same as SS\,433 bolometric luminosity (\cite{FaMe01}) one obtains
with the same opening angle of the funnel, the "observed" face-on luminosity 
of SS\,433 is $L_x \sim 10^{41}\,erg/s$ and the expected frequency of such objects
is $\sim 0.1$ per a galaxy like Milky Way.

The disk spherization radius (\cite{SS73}) in SS\,433 is estimated 
$r_{sp} \approx 3 \kappa \dot M_a/8 \upi c \sim 2.6 \cdot 10^{10}\,cm$, where
the Thomson opacity for a gas with solar abundance is $\kappa = 0.35$\,cm$^2$/g.
Corresponding wind velosity is $V_w \sim (G M / r_{sp}) \sim 2300 \,km/s$, where 
we adopted the black hole mass $M=10 \,M_{\odot}$. The size of the hot wind photosphere
and the photosphere temperature (for not a face-on observer) are estimated 
$r_{ph,w}= \dot M_{w} \kappa /4 \pi \cos \theta_f V_w \sim 1 \cdot 10^{12}\,cm$,  
$T_{ph,w}=(L_{bol}/ 4 \pi \sigma \cos \theta_f r_{ph, w}^2)^{1/4} \sim 6 \cdot10^4 \,K$.
Both the size of the hot body and the temperature are quite close to those 
we observe in SS\,433.

We find the jet photosphere size 
$r_{ph, j} = \dot M_j \kappa / \Omega_f V_j \sim 4 \cdot 10^9\,cm$,
this value indicates the bottom of the funnel, below $r_{ph, j}$ the funnel walls
can not be observed. A temperature of the inner funnel walls at a level of $r_{ph, j}$
is estimated as (\cite{FK06, FaKaAb06}) $T_{ph, f}$ between
$\sim 1.7 \cdot 10^7\,K$ and $\sim 1 \cdot 10^6\,K$. Such temperatures provide 
a high ionisation of elements needed for the line-locking mechanism (\cite{ShaMiRe86})
to operate for the jet acceleration. 

\begin{figure}
{\centering
 \includegraphics[height=6.6cm,angle=-90]{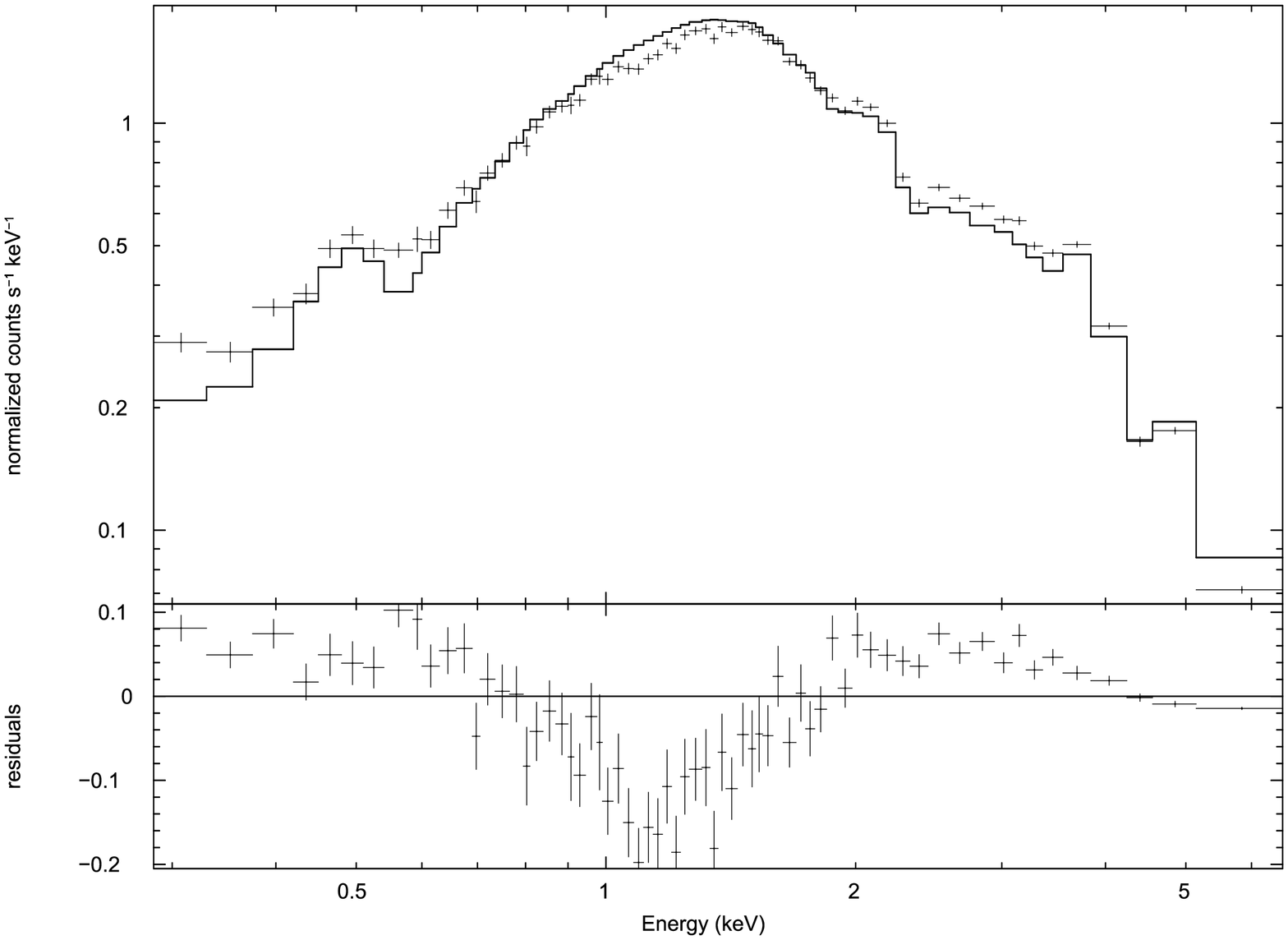}
 \includegraphics[height=6.6cm,angle=-90]{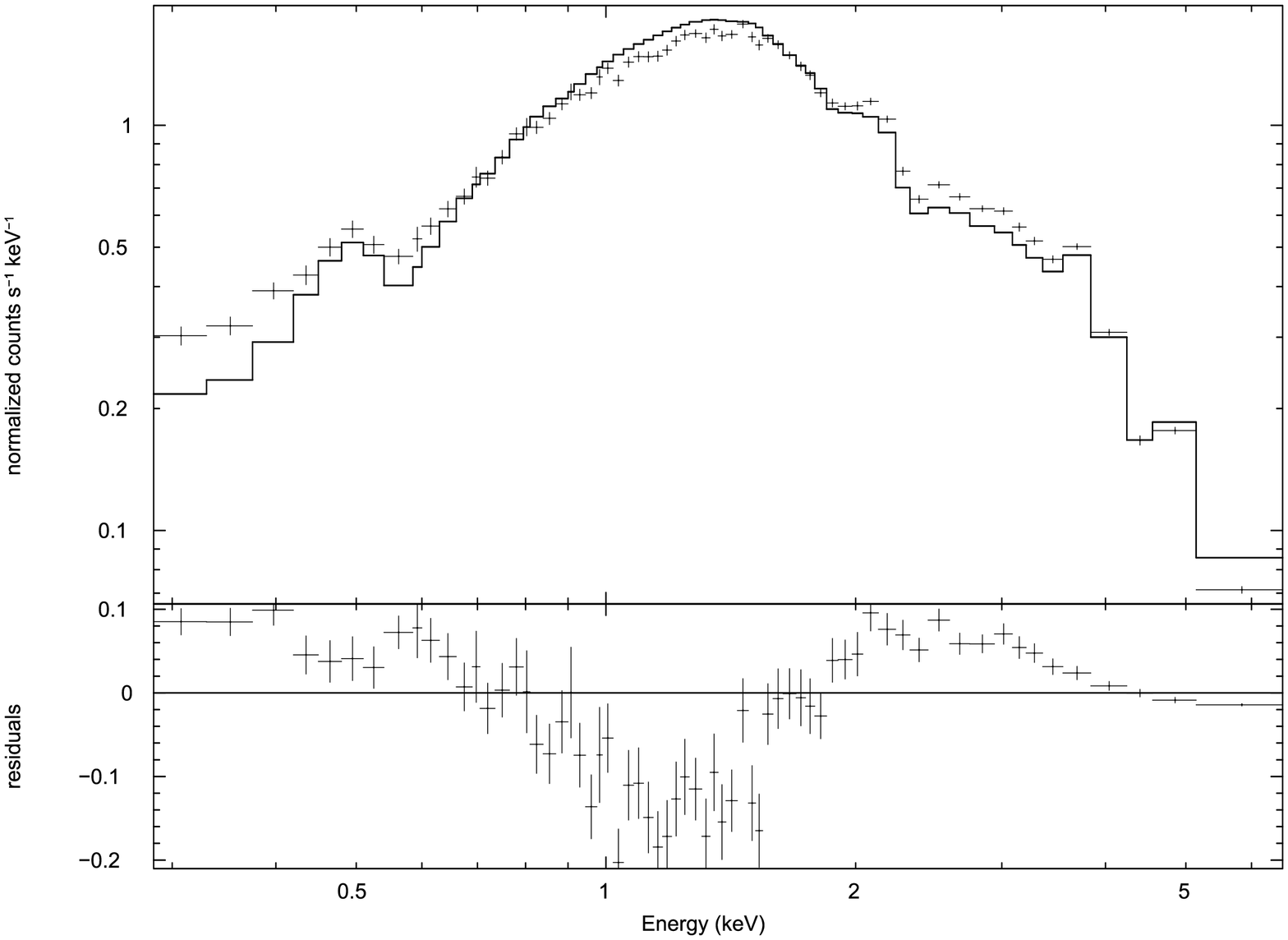}
\par}
  \caption{
Simulated MCF spectra as seen with XMM MOS1 with total counts $10^5$, 
obtained with $T_0 = 1\, keV$, $\theta_f = 20^{\circ}$, $V_w(r_0) = 6000\,km/s$, 
$N_H=0$ and $\xi=10^4$ (left), $\xi=10$ (right). Solid lines show the {\it diskbb} 
model fitted to the spectra. In the model spectra we found $T_{inn} = 1.55\, keV$, 
$N_H=1.5 \cdot 10^{21} \,cm^{-2}$.
The MCF spectrum is broader than the MCD (diskbb) spectrum.
}
\end{figure}

\begin{figure}
{\centering
 \includegraphics[height=6.2cm,angle=-90]{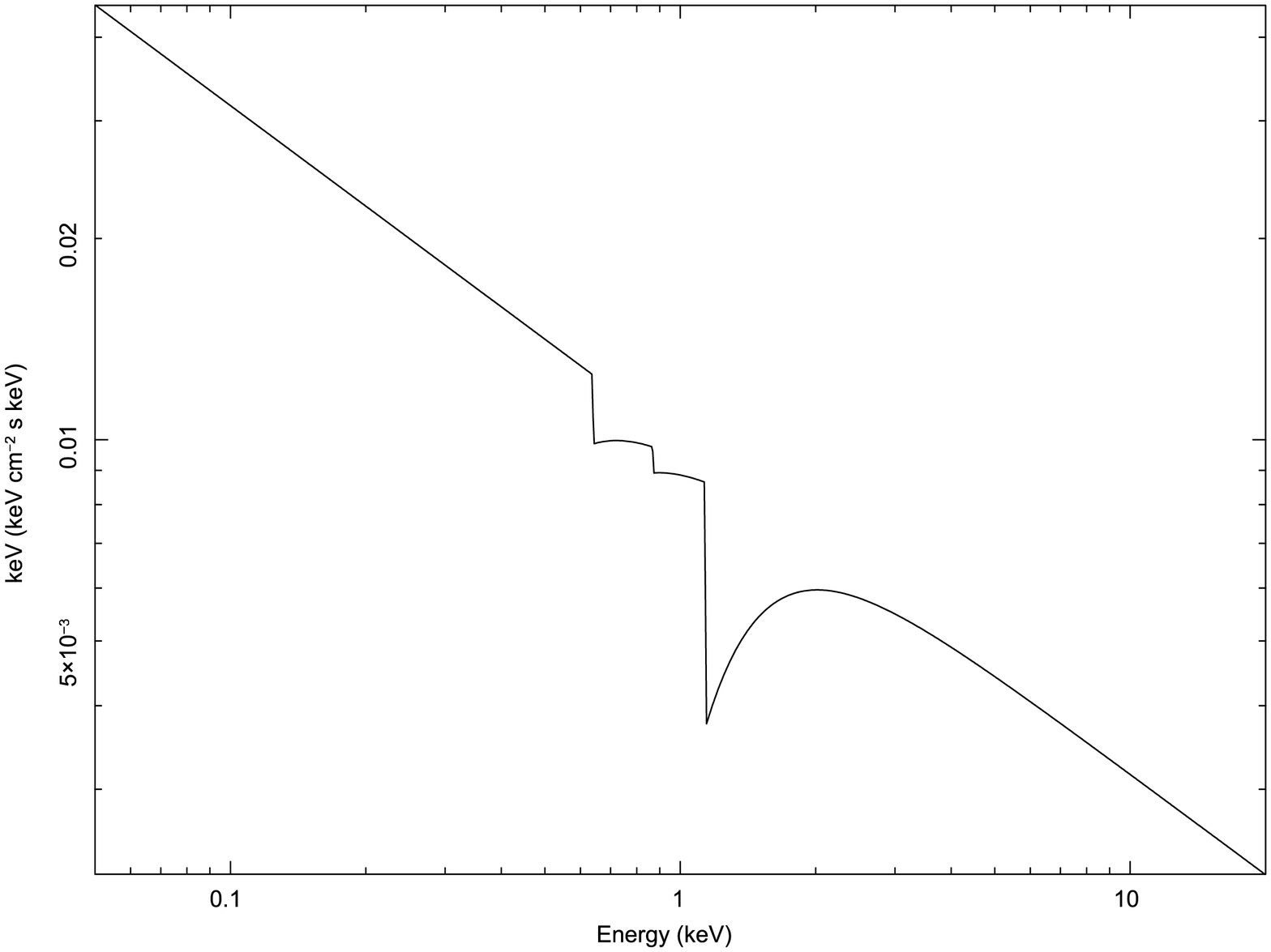}
 \includegraphics[height=6.5cm,angle=-90]{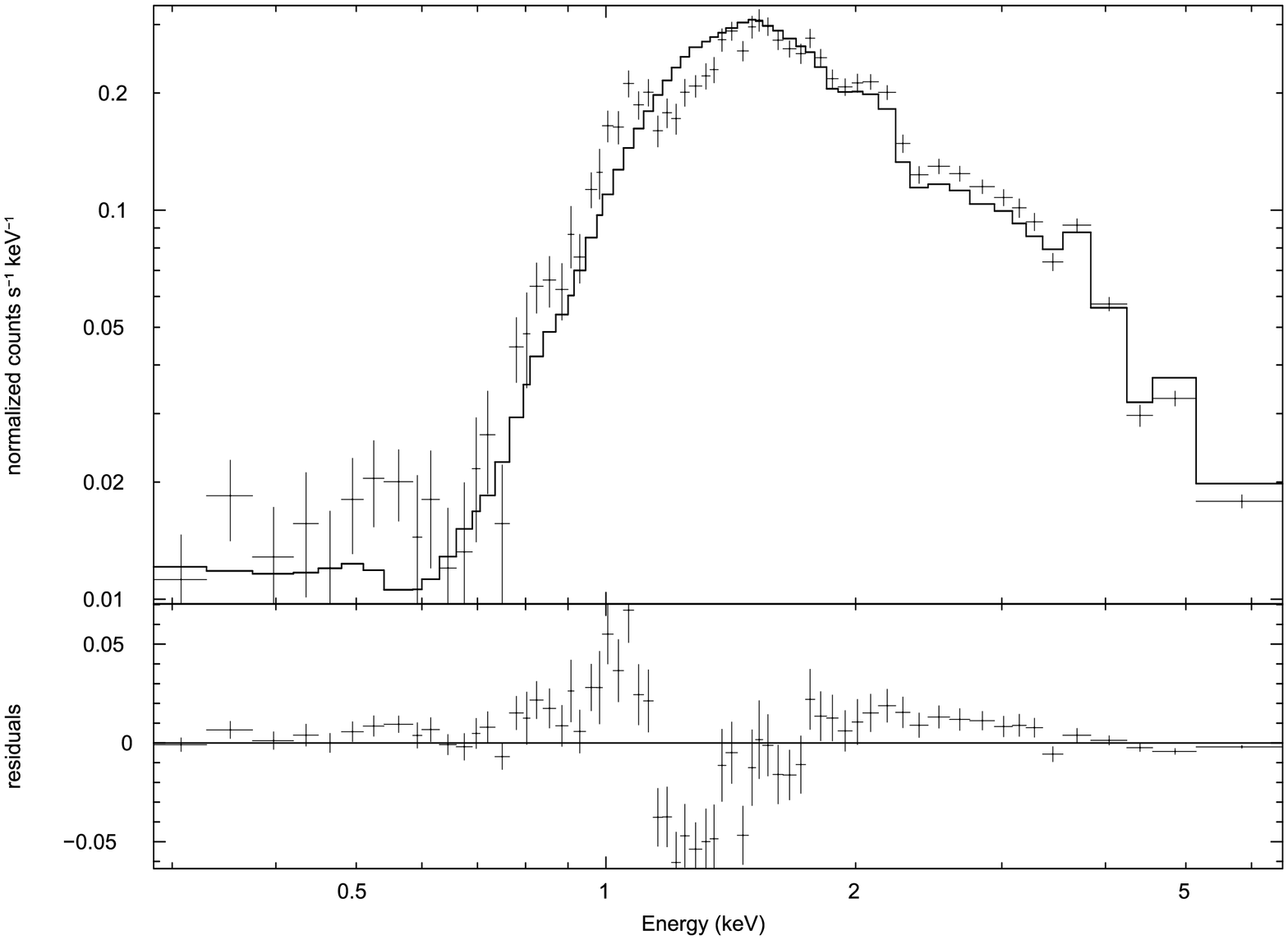}
\par}
  \caption{
Possible fake broad emission/absorption features in ULXs spectral residuals. 
The model is a power law spectrum ($\Gamma = 2.5$) with $N_H=1.0 \cdot 10^{22} \,cm^{-2}$
and with introduced Lc edges of C\,VI, N\,VII and O\,VIII blueshifted to the velocity 
$V_j=0.26 c$ (shown in left panel). The XMM MOS1 fit to the model 
($\Gamma = 2.2$, $N_H=1.05 \cdot 10^{22} \,cm^{-2}$) and spectral residuals 
are shown by solid line. 
}
\end{figure}

We conclude that the UV radiation of the SS\,433 disk ($T\sim 50000\,K, L\sim 10^{40} \,erg/s$) 
is roughly isotropic, but its X-ray funnel radiation 
($T\sim 10^6 - 10^7 \,K, L\sim 10^{40} \,erg/s$) 
is mildly anisotropic. The same conclusions have to be done for face-on SS\,433 stars (ULXs),
they are expected to be very bright UV sources and their X-ray luminosities have to be
even higher, up to $\sim 10^{40}-10^{41} \,erg/s$, due to the additional geometrical
collimation.  

Temporal variability of the SS\,433 funnel is expected on time scales 
$r_{ph, w}/c \sim 30\,sec$ and $r_{ph, j}/c \sim 0.1\, sec$. A typical accretion disk 
variability power density spectrum at scales $>> 0.1\, sec$ is expected. However the most
rapid variability may be observed only for face-on SS\,433 stars, in the case of 
SS\,433 (the edge-on system), all the short-scale variability of the funnel 
($\Delta t < 10-100 \,sec$) has to be smoothed down in the funnel. Some of the expected
X-ray variability patterns have been observed recently (\cite{Revn05}, \cite{Revn06, Kotani06}).

\section{The funnel spectrum and applications to ULXs}

The multicolor funnel (MCF) model has been developed (\cite{FK06, FaKaAb06}) 
to estimate the emerging X-ray spectrum in the funnel. The main parameters of 
the models are (i) the initial radius, the deepest visible level in the inner 
funnel walls ($r_0 = r_{ph, j}$), (ii) the walls temperature at $T_0(r_0)$, (iii) the
ratio of radiation to gas pressure $\xi={aT_0^3}/{3k_b n_0} $ at $r_0$, (iv)
the wind velocity $V_w(r_0)$ and (v) the funnel opening angle $\theta_f$.
Photon trapping in the wind is considered by a simple comparison of advection time
($t_{mov} \sim r/v(r)$) and diffusion time ($t_{esc} \sim \beta_t r_{ph,w}/V(r)$),
where $\beta_t$ is a terminal wind velocity. These times are equal at the radius
$r(t_{mov}=t_{esc}) = 2 \dot M_{w} r_{sp} / 3 \cos \theta_f \dot M_{a} \sim r_{sp}$.
There is a local advective radiation transfer at $r<<r_{sp}$ and global diffusion 
transfer at $r>>r_{sp}$. 

In Fig.\,1 we present simulated MCF spectra as seen with XMM MOS1 with total counts 
$10^5$, obtained with $T_0 = 1\, keV$, $\xi=10^4$ and 10, $\theta_f = 20^{\circ}$, 
$V_w(r_0) = 6000\,km/s$, $N_H=0$ and their fittings using the {\it diskbb} model.  
This figure demonstrates that the MCF spectrum is broader than that of the MCD (diskbb).

Shallow, very broad (0.1-0.3c) and blue-shifted absorption lines are 
expected in the funnel X-ray spectrum (\cite{Fab04, FK06}). The absorption bands 
should belong to H- and He-like ions of the most abundant (O, Ne, Mg, Si, S, Fe) 
elements and should extend from the Kc to the K$\alpha$ energies of the corresponding
ions. 

Fig.\,2 shows possible fake broad emission/absorption features which could be observed
in ULXs spectral residuals. We have simulated a power law spectrum with introduced
Lc edges of C\,VI, N\,VII and O\,VIII blueshifted to the SS\,433 jet velocity $V_j=0.26 c$.
The optical thickness of these introduced edges corresponds to "effective" hydrogen
thickness $\tau(L_c)=20$. Spectral residuals in the figure were derived from the XMM MOS1
power law spectral model to the simulated spectrum. The fake emission/absorption features
appear both due to the blueshifts of the edges and the energetic differences 
between neutral and highly ionised absorptions. Residuals of such type are observed in
ULXs spectra (for example, \cite{Dewan06}, \cite{Roberts06}). The absorption edges,
if they are confirmed in ULXs spectra give a possibility to measure the jet velocities
in supercritical disk funnels.

\begin{acknowledgments}
The work is supported by RFBR under grants number 03-02-16341, 04-02-16349
and by joint RFBR/JSPS grant N\,05-02-12710.
\end{acknowledgments}

\end{document}